\documentclass{article}

\usepackage{microtype}
\usepackage{graphicx}
\usepackage{subfigure}
\usepackage{booktabs} % for professional tables
\PassOptionsToPackage{hyphens}{url}

\newcommand{\R}{\mathbb R}

\usepackage{hyperref}

\usepackage[accepted]{icml2018}
\usepackage{amsmath,amssymb}

\icmltitlerunning{Collaboration of AI Agents via Cooperative Multi-Agent Deep Reinforcement Learning}

\begin{document}

\twocolumn[
\icmltitle{Collaboration of AI Agents via Cooperative Multi-Agent Deep Reinforcement Learning}

% It is OKAY to include author information, even for blind
% submissions: the style file will automatically remove it for you
% unless you've provided the [accepted] option to the icml2018
% package.

% List of affiliations: The first argument should be a (short)
% identifier you will use later to specify author affiliations
% Academic affiliations should list Department, University, City, Region, Country
% Industry affiliations should list Company, City, Region, Country

% You can specify symbols, otherwise they are numbered in order.
% Ideally, you should not use this facility. Affiliations will be numbered
% in order of appearance and this is the preferred way.
\icmlsetsymbol{equal}{*}

\begin{icmlauthorlist}
Niranjan Balachandar, Justin Dieter, Govardana Sachithanandam Ramachandran\\
*all authors contributed equally
\end{icmlauthorlist}

\vskip 0.3in
]

% this must go after the closing bracket ] following \twocolumn[ ...

% This command actually creates the footnote in the first column
% listing the affiliations and the copyright notice.
% The command takes one argument, which is text to display at the start of the footnote.
% The \icmlEqualContribution command is standard text for equal contribution.
% Remove it (just {}) if you do not need this facility.

\begin{abstract}
There are many AI tasks involving multiple interacting agents where agents should learn to cooperate and collaborate to effectively perform the task. Here we develop and evaluate various multi-agent protocols to train agents to collaborate with teammates in grid soccer. We train and evaluate our multi-agent methods against a team operating with a smart hand-coded policy. As a baseline, we train agents concurrently and independently, with no communication. Our collaborative protocols were parameter sharing, coordinated learning with communication, and counterfactual policy gradients. Against the hand-coded team, the team trained with parameter sharing and the team trained with coordinated learning performed the best, scoring on 89.5\% and 94.5\%  of episodes respectively when playing against the hand-coded team. Against the parameter sharing team, with adversarial training the coordinated learning team scored on 75\% of the episodes, indicating it is the most adaptable of our methods. The insights gained from our work can be applied to other domains where multi-agent collaboration could be beneficial.    
\end{abstract}

\section{Introduction}	
There are numerous tasks where collaboration is beneficial: driving, multi-player games, etc. In the future, many of the cars on the road will be autonomous \cite{autonomous}, so it would be useful for these cars to learn to collaborate to avoid accidents and improve efficiency. In multi-player games, it would be beneficial for agents to operate under a coordinated protocol that leverages unified strategies with specialization of responsibilities. A good example of such a game is soccer, where players have a common team strategy while performing individual roles. 

While learning to collaborate for these tasks seems intuitive to humans, for a given task, it is unclear how to most effectively balance AI collaboration with specialization. We might consider treating the entire multi-agent system as a single agent and train this agent with a centralized controller. However, for a centralized controller the state and action spaces rise exponentially with the number of sub-agents, so this would not be logistically practical or computationally tractable when the state-space is large or there are many agents. 

Thus, decentralized methods that are capable of allowing collaboration and communication between agents are desirable for these situations. Ideally, the policies learned by the agents via decentralized methods will be adaptable to variations in the policies of collaborating agents, generalizable, and capable of achieving similar performance to a that of a centralized controller. Here we explore and evaluate various multi-agent reinforcement learning methods in application to grid soccer, a multi-player soccer game \cite{gridsoccer}, our objective is to train our multi-agent reinforcement learning teams to defeat a team operating with an intelligent handed-coded strategy. 

\section{Background/Related Work}
A number of methods for multi-agent control have been developed in literature.

\subsection{Centralized} 
This approach involves a centralized controller that maps states of all the agents to a joint action (an action for each agent) \cite{multimethods}. This is a Multi-agent Partially observable Markov decision process (MPOMDP) policy. 

\textbf{Advantages:} This scheme allows tighter coordination between agents and the dynamics are stationary. 

\textbf{Disadvantages:} The main disadvantage of the centralized approach is the exponential increase in the state and action spaces with increase in number of agents.  Manipulations can be performed to reduce the joint action space size to $n|A|$ \cite{central} for a discrete action space. For example, we could reduce the action space by factoring action probability as $P(\vec{a})=\prod_i P(a_i)$ where $a_i$ are the individual actions of agents, which reduces the action space from $|A|^n$ to $n|A|$. However, when there are many collaborating agents, this method may still be impractical. Thus for our paper, we will focus on decentralized methods for multi-agent control.

\subsection{Concurrent Learning}
In concurrent learning, each agent is independent and learns its own unique policy \cite{multimethods}. In this scheme each agent has an independent observation and independently executes actions based on its observation. Here the reward is shared between all the agents. Agents should learn their individual roles while being evaluated with team performance, so this method should effectively decompose coordination in a distributed way. 

\textbf{Advantages:} The agents learn heterogeneous policies and the agents take specific roles, for example, in the soccer simulator, an agent could learn to be defender, another can learn to be the striker. Another advantage is the there is no communication between agents, which may be useful in a system where there are bandwidth constraints.

\textbf{Disadvantages:} Since the agents do not share their experience with each other, the training of unique policies does not scale with a large number of agents. This adds to the sample complexity of the scheme. Another disadvantage is that since all agents are learning policies independently, from each agent's point of view the dynamics are not stationary.

% \subsection{Parameter Sharing} brief description

\subsection{State sharing} 
In this scheme, neighboring agents share their observation and action space with each other \cite{statesharing}. Here the neighboring agents' actions and observation are stacked as the state space of each agent. This reduces the problem to distributed constraint optimization (DCOP).   

\subsection{Counterfactual Policy Gradients}
Counterfactual policy gradients is an actor critic method that leverages the utility of a centralized value approximation network as with parameter sharing but also gives a method for credit assignment to ensure that agents are properly given credit for individual actions rather than the actions of the group and allows for specialization of individual agents \cite{counterfactual}	. The agents are represented by decentralized policy networks.

The counterfactual algorithm uses TD learning to update the centralized critic and gradients of the policy network to perform updates. The policy gradients use a specialized advantage function that analyzes the contributions of each agents individual actions to the overall received reward.

\section{Approach}

\subsection{Grid Soccer Simulator}
\begin{figure*}[hbpt!]
\centering
\includegraphics[width=0.7\textwidth]{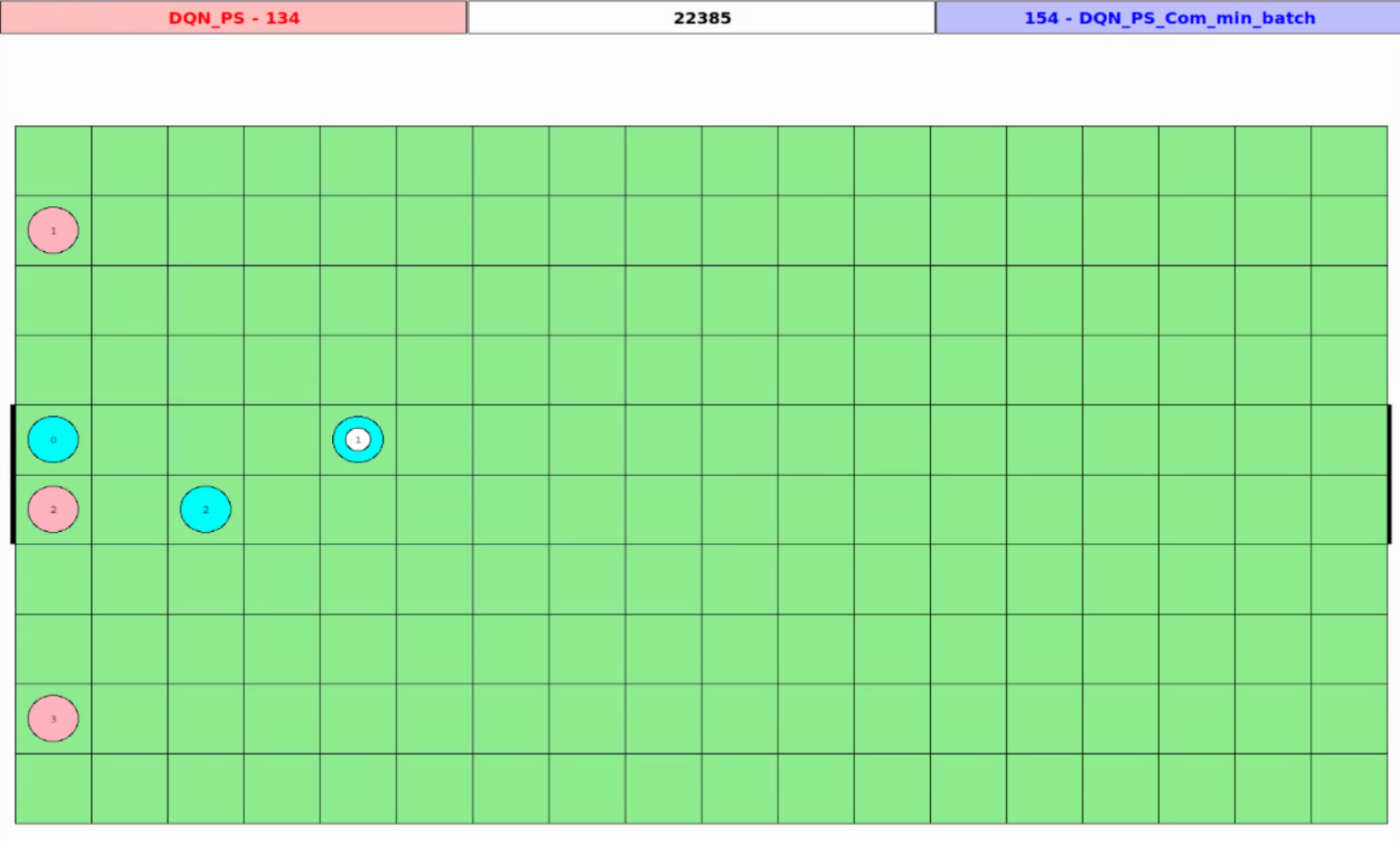}
\caption{Sample rendering of grid soccer game state. Two of our AI teams are playing against each other.}
\label{gridsoccer}
\end{figure*}

Grid Soccer Simulator is a discrete grid simulation of multi-player soccer \cite{gridsoccer}. We built a python API to interact with this grid soccer simulator. The two teams have 3 players each that play on a grid of size $10\times 18$, as shown in Figure \ref{gridsoccer}. Goals are indicated by the bolded lines on the right and left sides of the grids, and at any given timestep, one of the 6 players has the ball. 

In order to score, a player must physically move to the goal. It is possible (but obviously not advisable) for a player to score in its own goal, thus rewarding the opposing team. At each timestep, players can hold (stay still), move to one of the 8 adjacent grid positions, or pass to a teammate. choosing to move to an invalid position causes a player to remain at the same position. After each goal, a player from the team that did not score starts with the ball. Players can steal (cause a turnover of the ball) by running into a player of the opposing team who has the ball and is not holding or standing in between teammates of the opposing team that are passing to each other. 

The grid soccer game comes with a team that operates with an intelligent hand-coded strategy that is based on real-world soccer strategies. For example, when the opposing team has the ball and is close to the hand-coded team's goal, the players on the hand-coded team will go towards the hand-coded team's goal to play defense.

\subsection{States}
\begin{figure}[hbpt!]
\centering
\includegraphics[width=0.4\columnwidth]{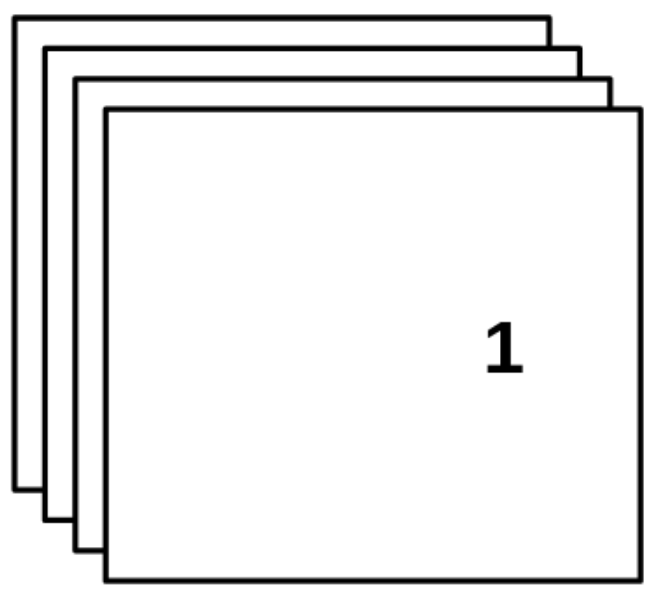}
\caption{illustration of state representation as boolean 4-channel images}
\label{state}
\end{figure}

We intend to leverage the ability of Convolutional Neural Networks to analyze images, so at each timestep, we represent the state of each agent as a 4-channel image $S \in \R^{H \times W \times 4}$ where $H$ and $W$ are the height and width of the grid respectively as illustrated in Figure \ref{state}.  Each pixel of $S$ is a boolean indicating the presence of a certain object. The first channel indicates the agent location:
$$S_{i,j,1} = \begin{cases} 
      1 & \text{location of agent is }(i,j) \\
      0 & \text{otherwise}
   \end{cases}.
$$
The second channel indicates the teammates' locations:
$$S_{i,j,2} = \begin{cases} 
      1 & \text{location of any teammate is }(i,j) \\
      0 & \text{otherwise}
   \end{cases}.
$$
The third channel indicates the opposing team members' locations:
$$S_{i,j,3} = \begin{cases} 
      1 & \text{location of any opposing team member is }(i,j) \\
      0 & \text{otherwise}
   \end{cases}.
$$
The fourth channel indicates the ball's location:
$$S_{i,j,4} = \begin{cases} 
      1 & \text{location of ball is }(i,j) \\
      0 & \text{otherwise}
   \end{cases}.
$$
\subsection{Actions}
There are a total of $n+8$ actions, where $n$ is the number of players per team ($n=3$ for our experiments). The possible actions for an agent currently located at $(i,j)$ are listed in Table \ref{actions}.
\begin{table}[hbpt!]
\centering

 \begin{tabular}{| l |l |}
 \hline
 Action Index & Action \\
 \hline
 \hline
 0 & stay at (i,j) \\  \hline
 1 & move to (i+1,j) \\  \hline 
 2 & move to (i,j-1) \\  \hline 
 3 & move to (i-1,j) \\  \hline 
 4 & move to (i,j+1) \\  \hline 
 5 & move to (i+1,j+1) \\  \hline 
 6 & move to (i+1,j-1) \\  \hline 
 7 & move to (i-1,j-1) \\  \hline 
 8 & move to (i-1,j+1) \\  \hline 
 $8+k$ for $k=1$ to $n-1$ & pass to teammate $k$ \\  \hline

 \end{tabular}
 \caption{possible actions for each agent if current agent location is $(i,j)$ and each team has $n$ players}
 \label{actions}
\end{table}

\subsection{Rewards}
We constructed a handcrafted reward function to penalize getting scored on, turnovers, and stalling to incentive stealing the ball and scoring quickly. At each timestep, all agents in the team receive a reward, which depends on the result of the timestep. The possible results of a timestep and the corresponding rewards are summarized in Table \ref{rewards}. 
\begin{table}[hbpt!]
\centering

 \begin{tabular}{| l |r |}
 \hline
 Timestep Result & Reward \\
 \hline
 \hline
 agent own-goal & -100 \\  \hline
 team own-goal & -75 \\  \hline 
 agent scored goal & 50 \\  \hline 
 team scored goal & 50 \\  \hline 
 opponent scored goal & -50 \\  \hline 
 opponent own goal & 10 \\  \hline 
 agent turns ball over & -10 \\  \hline 
 team turns ball over & -10 \\  \hline 
 agent steals ball & 10 \\  \hline 
 team steals ball & 10 \\  \hline 
 agent illegal movement & -3\\  \hline 
 agent successful pass & -1\\  \hline 
 agent hold & -1\\  \hline 
 agent legal movement & -2\\  \hline

 \end{tabular}
 \caption{Rewards for each possible timestep result}
 \label{rewards}
\end{table}

\subsection{Deep Q-Learning}
\begin{figure*}[hbpt!]
\centering
\includegraphics[width=\textwidth]{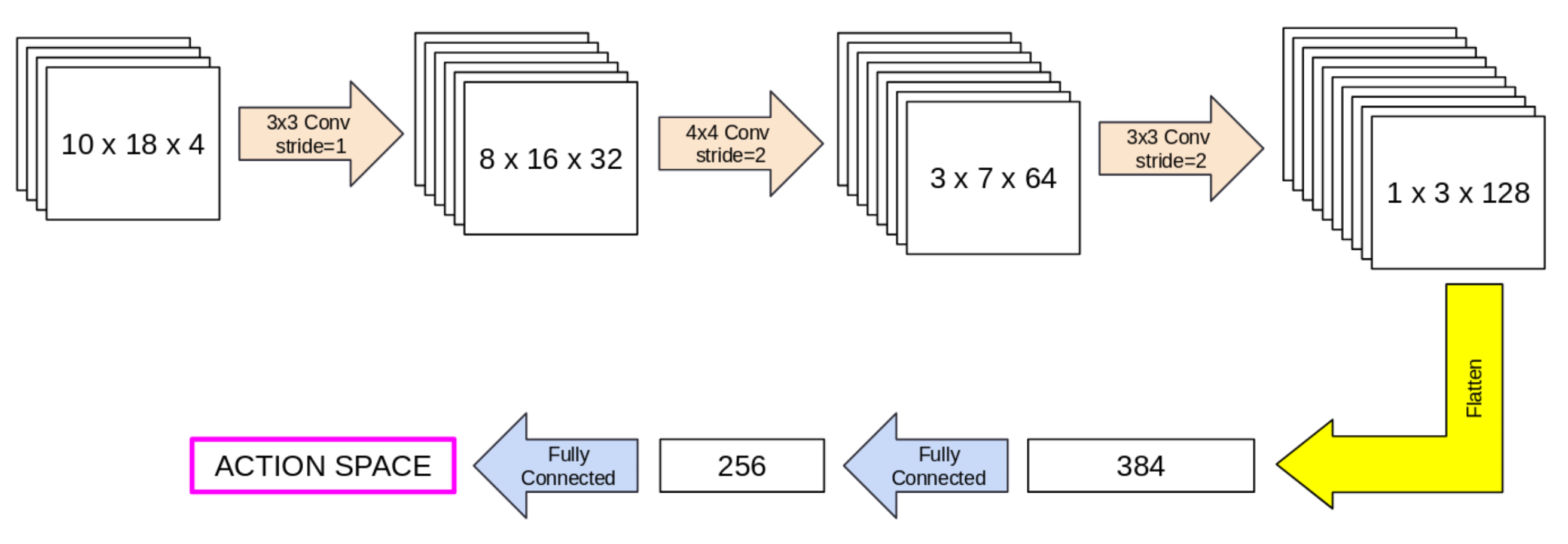}
\caption{Q-value approximation network architecture}
\label{cnn}
\end{figure*}

Because the state space is very large (around $(2n)^{HW+1} = 6^{181}$ for $n=3$ players per team, tabular Q-learning is impractical, so we approximate the Q-value with a Convolutional Neural Network (CNN). Our CNN takes as input an agent state ($H\times W \times 4$) image and outputs a vector of size $|A|$ of Q-values for each action for the input state. Our CNN has 3 convolutional layers followed by 2 fully connected layers, and the architecture is summarized in Figure \ref{cnn}.  Each layer except the final fully connected layer has ReLU activation. For stability, we include a target network as included \cite{dqn}. We update the target network weights update after every iteration in which a goal is scored. We optimize the following objective function for gradient descent:
$$L(\theta) = \left(r + \gamma \max_{a'} \hat{Q}(s',a',\theta^-) - \hat{Q}(s,a,\theta)\right)^2$$
where $\theta$ are the network weights and $\theta^-$ are the target network weights, and $\hat{Q}$ is the feed-forward output of the CNN. During training, we perform an epsilon-greedy exploration with $\epsilon$ decayed over time and gradient descent with Adam \cite{adam}.

\subsection{Concurrent Learning}
Our baseline approach is concurrent learning with no communication. Here each agent learns its own set Q-network weights. Since this approach has no explicit communication or collaboration protocols, we expect this approach to perform the poorest.

\subsection{Parameter Sharing}
In the parameter sharing scheme, the agents are homogeneous. This allows model to be trained with experiences of all agents, which makes the algorithm data efficient. Here the learning is centralized while the control is decentralized by sharing learned parameters of the model between the agents. The agents still have different behavior since each agent receive different observations. This means a single policy is learned during train time and each agent executes the policy during inference with their independent observation.

Advantage: The experience of all agents are used by the learner, hence making them sample efficient. Since the model is learned by a centralized learner and shared across all agents, they are  computationally efficient. Computational resources such as GPU can be allocated to the centralized learner and each agent should use only light computational resources since inference is not computationally intensive.
Disadvantage: It requires a centralized leaner and all of the agent's observation, reward and action needs to be  transmitted to the leaner. This might be a limitation for an application that has limited bandwidth. The model lacks tighter coordination between agents as there is no possibility of communication. Here the model learns co-ordination implicitly by learning protocols an individual agent should follow based solely on its state.

Here we use Q-Learning to update our value function approximations. State, action, reward and the next sate from agents are transmitted to the centralized learner. For each time step of each agent, the centralized learner computes the Q-learning gradient to update the Q network. All agents share the same Q-network weights.

\subsection{Coordinated Learning with Communication}

The main limitation of parameter sharing is that they lack tight co-ordination between agents. We address this by defining a communication mechanism between agents. Many approaches in the past that use communication for tighter co-ordination, formulate the problem as Distributed Constraint optimization (DCOP), which involves usage of the whole system or at least a smaller subsystem of agents. This uses significant communication bandwidth and causes latency in action selection for each agent. This becomes more intractable as the number of agents in the system increases.

We formulate a new form of communication that uses a very small amount of bandwidth where the action selection is decentralized and  does not involve computation for a group of agents. We achieved this by defining a joint action with communication $a^*$:
$$a^* = (a_i, a_g)$$
here $a_i$ is the action taken by an individual agent. $a_g$ is one of a discrete number of communications that is broadcasted to other agents. The communications from neighboring agents are appended to an agents own observation $x_{it}$, to form the state space
$$s_{it} = x_{it},a_{jt} \forall j, j\neq i$$
In our DQN, we allow agents to receive this communication by having the following revised agent state representation. Instead of the original 4 channels, there will now be $3+|A_g|$ channels. The agent location channel, opponents location channel, and ball location channel remain. But now instead of a teammates locations channel, there will be $|A_g|$ channels, where channel $a_g$ will have a 1 at pixel $(i,j)$ if there is a teammate at $(i,j)$ that broadcasted communication action $a_g$ at the most recent timestep. Now our Q-network will output to joint action space instead of action space. Our implementation of coordinated learning also includes parameter sharing (each agent has the same Q-network weights) for greater sample efficiency.

One of the limitations of this approach is that the action space increases to $|a_i|\times|a_g|$. Hence discretion is required in choosing the number of available group actions. In our experiments on grid soccer, we achieve this by defining a joint action:
$$a^* = \arg \max_{a_t}Q(s_t, a_t)$$
Another limitation is that since each agent is learning to communicate the group actions and since the neighboring agents' group actions are used as part of this agent's state space, the dynamics of the model are not stationary. To mitigate this issue, we use minibatches and a replay buffer for the model updates. For each time step for each agent, we capture experience $e_t=(s_t,a_t,r_t,s_{t+1})$ and we update a replay buffer $R={e_1,e_2,...e_n}$. At each epoch, we sample a minibatch from the replay buffer $R$. Use of minibatch and replay buffer improves stability.

Though coordinated learning with communication performs best, it requires a lot of epochs to train. One of the main reasons for this is because, each agents receives the same reward. We address this by using appropriate credit assignment. The approach we explored is difference reward \cite{approx}, which is defined as
$$D_i(z) = G(z) - G(z_i+c_i)$$
Here $G(z)$ is the global reward of unified state $z$ of the system, $z_i$ is the system state without agent $i$, and $c_i$ is the counterfactual term. Since it is non-trivial to estimate $G(z_i+c_i)$, we formulated a simple strategy for credit assignment. We assign a reward for each agent based on the following:
$$R_i= R . Q_i(s_t,a_t) (\frac{1}{N} \sum_{i=1}^{N} Q_i(s_{it},a_{it}))$$
Here $N$ is number of agents in the system, $R$ is global reward signal, and $Q_i(s_{it},a_{it})$ is the state action value for agent $i$ at timestep $t$. The intuition behind this is that if the agent directly contributed to the reward, the corresponding $Q_i$ would be greater than for an agent which did not contribute to overall performance of the system. 

\subsection{Counterfactual Policy Gradients}
The implementation of the counterfactual actor critic policy gradient method works as follows:

\subsubsection{Centralized Critic}
\begin{figure}[hbpt!]
\centering
\includegraphics[width=0.6\columnwidth]{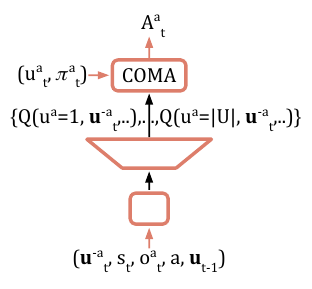}
\caption{Structure of the centralized critic for the counterfactual algorithm. $\mathbf{u}_t$ denotes the joint action at timestep $t$ and $\mathbf{u}^{-a}_t$ denotes the joint action excluding the action of agent $a$ \cite{counterfactual}}
\label{comacritic}
\end{figure}

The centralized critic is a function approximator
\[
Q: S \times \mathbf{A} \rightarrow \mathbb{R}
\]
where $\mathbf{A}$ is the joint action space of all agents. The usual representation of value approximation with a neural network that outputs a vector of size $|\mathbf{A}|$ will not work as $|\mathbf{A}|$ will grow exponentially with the number of agents and will become infeasible to represent as the output of a neural network. Instead, the centralized critic can output a vector of size $|A|$ to provide the $Q$ values of a single agent where $A$ is the action space of a single agent. As an input, the network takes in the other agents actions as well as a one-hot vector representing which agent $Q$-values are being produced for. This network structure is demonstrated in Figure \ref{comacritic}.

The details of the actual model implemented are as follows. The states were represented as a 3 dimensional tensor, with the first two dimensions the dimensions of the grid followed by one channel each for every player on the counterfactual team, a single channel for every player on the opposing team, and a channel for the location of the ball. This was fed into a convolutional layer with $32$ output channels, a $3 \times 3$ kernel, and a stride of $1$. This was then fed into another convolutional layer with $64$ output channels, a $4 \times 4$ kernel, and a stride of $2$. This is then flattened and a one hot representation of $\mathbf{u}^{-a}$ as well as a one hot representation of $a$ is appended before being processed by two fully connected layers with ReLu activation and respective sizes of $128$ and $64$. Finally a linear layer is added at the end to output a vector of size $|A|$. The network is trained using the SARSA($\lambda$) algorithm which gives a combination of SARSA style returns and Monte Carlo returns \cite{counterfactual}. We used $\lambda = 0.8$ to be empirically effective but the model could potentially benefit from more fine tuning of this parameter.

\subsubsection{Policy Network}
To represent the stochastic policies of the actors, a policy network with a softmax output of dimension $|A|$ is used. To allow the agents to share what they have learned parameters are shared but a one-hot representation of which agent is using the network is passed as an input to enable specialization of agents. The exact structure of the neural network is as follows. The state is represented as a 3 dimensional tensor with a channel for the current agent's locations, a channel for the ball's location, a channel for the location of teammates, and a channel for the location of opposing players. This is fed into a convolutional layer with $32$ output channels, a $3 \times 3$ kernel, and a stride of $1$. This was then fed into another convolutional layer with $64$ output channels, a $4 \times 4$ kernel, and a stride of $2$. This is then flattened and a one hot representation of $a$ is appended before being processed by two fully connected layers with ReLu activation and respective sizes of $128$ and $64$. A softmax layer is added at the end to give an output dimension of $|A|$. Additionally a decaying randomness factor of $\epsilon$ is added such that $\pi(a | s) = (1 - \epsilon) \pi(a | s) + \epsilon$ to ensure that the agent is exploring sufficiently. $\epsilon$ is decayed from $0.5$ to 
$0.05$ over $300000$ timesteps.

\subsubsection{Actor Training and Counterfactual Advantages}
The unique representation of the centralized critic allows for a very useful advantage function to be computed.
\[
A^{a}(s,\mathbf{u}) = Q(s, \mathbf{u}) - \sum_{u'^a}\pi^{a}(u'^{a} | \tau^{a})Q(s,(\mathbf{u}^{-a}, u'^{a}))
\]
This counterfactual advantage provides a metric of how much better/worse an agents actions will be compared to if it had chosen a different action with the actions of other agents held fixed. This provides a method of credit assignment to ensure that agents are only rewarded for how their actions contribute to the success of the team verses getting rewarded/penalized for results they did not contribute to. This advantage is then fed into the typical policy gradient update rule with advantages:
\[
\Delta \theta = \alpha \sum_a \nabla_\theta log \pi^{a} (u^a | \tau^a) A^a (s, \mathbf{u})
\]
where $\theta$ is the parameters of the policy network. $\alpha$ is the learning rate, empirically chosen to be $0.001$ for our experiments. Also, all the theoretical guarantees of convergence with normal actor critic methods hold for this actor critic method with counterfactual advantages \cite{counterfactual}.

\section{Experiment Results}
The current code for all our simulators, APIs, and learning algorithms is included in \url{https://github.com/jdietz31/CS234-MultiagentProject}. We evaluate the performance of our algorithms by measuring the ``goal ratio'', which is the proportion of goals scored by the algorithm's team of the last 200 goals. So the worst possible goal ratio an algorithm can achieve is $0$, and the best possible goal ratio an algorithm can achieve is $1$. We train each of our methods for hundreds of thousands of iterations. For each of our models, we use a learning rate of $0.001$, initial epsilon of $0.5$, final epsilon of $0.05$, and perform linear epsilon decay over the course of training. We vary the decay rate to account for differing sample efficiencies. For coordinated learning, we set the minibatch size to $1,000$ and replay buffer size to $50,000$. Hyperparameters for Counterfactual are specified in the methods section.
\begin{figure*}[htp!]
\centering
\includegraphics[width=1\textwidth]{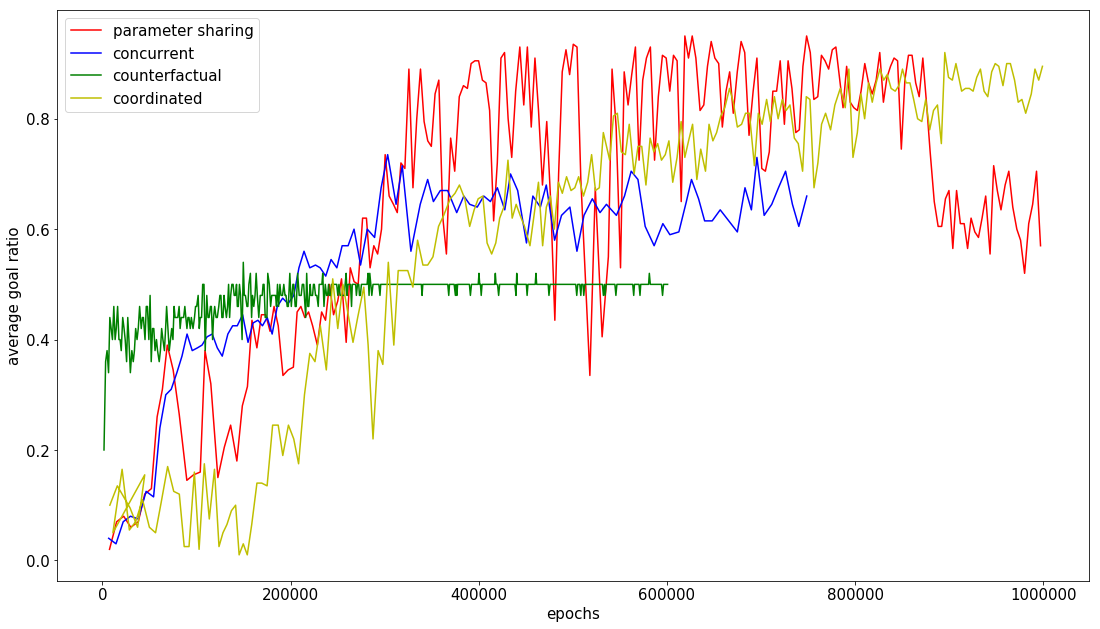}
\caption{Performance of multiagent models throughout training in terms of ratio of goals scored against the hand-coded team.}
\label{curve}
\end{figure*}

\begin{figure*}[htp!]
\centering
\includegraphics[width=0.8\textwidth]{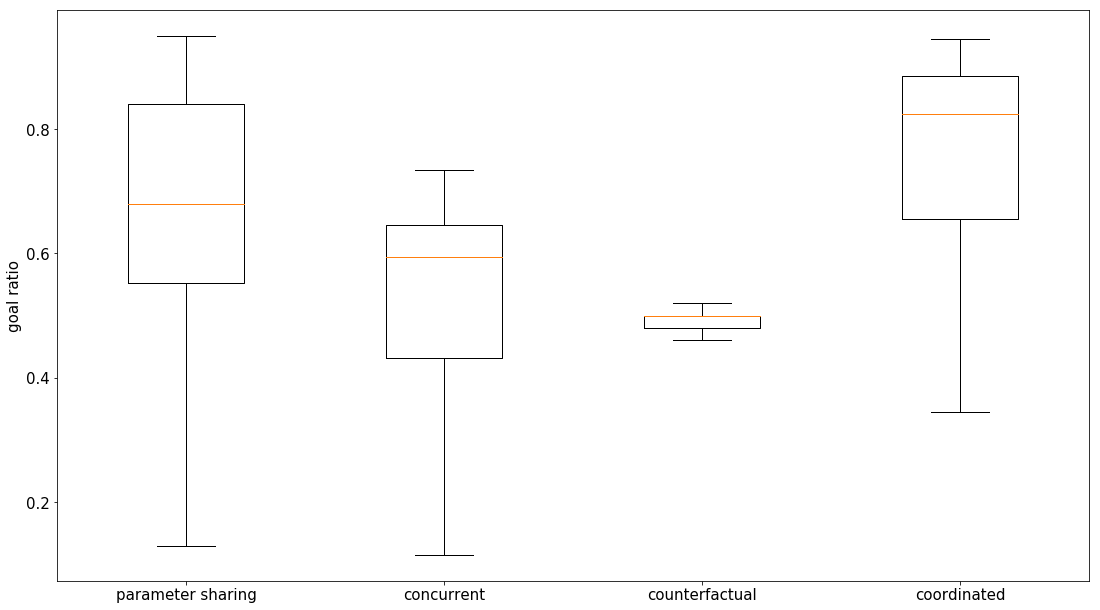}
\caption{Performance of multiagent models in the last 1000 iterations of training.}
\label{bar}
\end{figure*}

\begin{figure*}[htp!]
\centering
\includegraphics[width=0.8\textwidth]{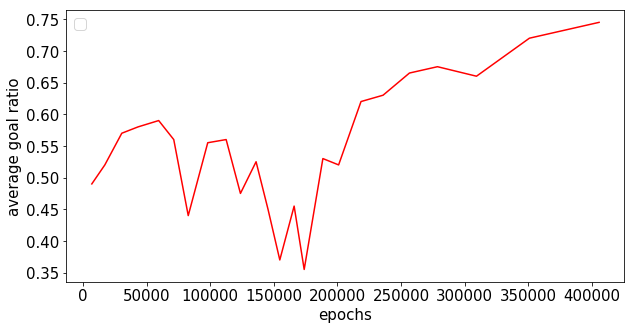}
\caption{Goal ratio of the coordinated learning team when training adversarially against parameter sharing after both are pretrained against the hand-coded team.}
\label{adversarial}
\end{figure*}

We then trained our two best models- parameter sharing vs coordinated learning against each other to further delineate performance. The results of this adversarial training are shown in Figure \ref{adversarial}. Videos of the parameter sharing team playing the coordinated learning team are included here: \url{https://drive.google.com/open?id=1jntEENSbnW_oyoFZJuCobBHybO_jdpUy}. 

The results in Figure \ref{curve} and Figure \ref{bar} shows the training performance and final performance of all of our different multi-agent models in terms of the ratio of goals scored against the hand-coded agent. The Concurrent model achieves decent performance, but we observe that one of the agents tends to learn offense and defense, while the other two learn very little, so there is almost no teamwork. Counterfactual learns quickly but converges to a local optima where it only plays offense and no defense. Parameter Sharing learns to be almost perfect but is slightly unstable in its performance throughout training. The coordinated model learns the slowest but is stable and converges to near perfect performance. Also, when allowed to train adversarially against other models after being pretrained against the hand-coded team, coordinated learning performs the best as demonstrated in Figure \ref{adversarial}.

\section{Conclusion}
Our results show that the model with communication is able to perform the best against both the hand coded agent and against our other reinforcement learning agents. It is a simple enough model that it is able to learn very effectively and its communication model allows it to learn methods of cooperating and team strategy that simpler models cannot compete with. The counterfactual policy gradient method is the most versatile and complex model and with sufficient hyper-parameter tuning it should be able to surpass other models. The complexity of the model makes it promising to be able to learn difficult tasks but also makes it very unstable and not very robust with respect to performing well with a variety of hyper-parameters. Future work can expand on making this counterfactual method more robust and less unstable. Also, our experiments only worked with 3 agents and real world multiagent systems can be composed of many more agents. Future work should investigate into how these models can perform in more complicated scenarios with many more agents.

\bibliography{example_paper}
\bibliographystyle{icml2018}

\end{document}